\documentclass[twocolumn,tighten]{aastex63}
\usepackage{xcolor}
\usepackage{graphbox}
\newcommand{\unit}[1]{\, {\rm #1}}
\graphicspath{{./}{figures/}}

\received{August 20, 2019}
\revised{February 6, 2020}
\accepted{February 13, 2020}
\submitjournal{ApJ}

\shorttitle{RSG Wind from the Progenitor of Cas~A}
\shortauthors{Weil et al.}

\begin{document}
\title{Detection of the Red Supergiant Wind from the Progenitor of Cassiopeia A}

\author[0000-0002-8360-0831]{Kathryn E.\ Weil}
\affil{Department of Physics and Astronomy, 6127 Wilder Laboratory, Dartmouth College, Hanover, NH 03755, USA}
\affil{Smithsonian Astrophysical Observatory, 60 Garden Street, Cambridge, MA 02138, USA}
\correspondingauthor{Kathryn E.\ Weil}
\email{kathryn.e.weil.gr@dartmouth.edu}

\author[0000-0003-3829-2056]{Robert A.\ Fesen}
\affil{Department of Physics and Astronomy, 6127 Wilder Laboratory, Dartmouth College, Hanover, NH 03755, USA}

\author[0000-0002-7507-8115]{Daniel J.\ Patnaude}
\affil{Smithsonian Astrophysical Observatory, 60 Garden Street, Cambridge, MA 02138, USA}

\author[0000-0002-7868-1622]{John C.\ Raymond}
\affiliation{Harvard-Smithsonian Center for Astrophysics, 60 Garden St., Cambridge, MA 02138, USA}

\author[0000-0002-9117-7244]{Roger A.\ Chevalier}
\affiliation{Department of Astronomy, University of Virginia, 530 McCormick Road, Charlottesville, VA 22904, USA}

\author[0000-0002-0763-3885]{Dan Milisavljevic}
\affil{Department of Physics and Astronomy, Purdue University, 525 Northwestern Avenue, West Lafayette, IN 47907, USA}

\author{Christopher L.\ Gerardy}
\affil{Department of Physics, University of North Carolina, Charlotte, NC 28223, USA}

\begin{abstract}
Cassiopeia A (Cas~A) is one of the best studied young Galactic supernova remnants. While providing a rare opportunity to study in detail the remnant of a Type IIb supernova, questions remain regarding the nature of its progenitor, its mass-loss history, and its pre-SN evolution. Here we present an optical investigation of the circumstellar environment around Cas~A and find clumpy and filamentary H$\alpha$ emission nebulosities concentrated 10--15\,pc (10--15 arcminutes) to the north and east. First reported by Minkowski as a faint H II region, these nebulosities exhibit distinct morphological and spectroscopic properties relative to the surrounding diffuse emissions. Compared to neighboring H~II regions, these nebulae show stronger [\ion{N}{2}] 6548, 6583\,\AA\ and [\ion{S}{2}] 6716, 6731\,\AA\ emissions relative to H$\alpha$. We show that Cas~A's highest-velocity ejecta knots are interacting with some of the closest projected emission nebulae, thus providing strong evidence that these nebulae lie at the same distance as the remnant. We interpret these surrounding nebulosities to be the remains of the progenitor's red supergiant wind which accumulated against the southern edge of a large extended H~II region located north of Cas~A. Our findings are consistent with the view that Cas~A's progenitor underwent considerable mass-loss, first from a fast main-sequence wind, then from a slower, clumpy red supergiant wind, and finally from a brief high-velocity wind, like that from a yellow supergiant.
\end{abstract}
\keywords{ISM: individual objects (Cassiopeia A), ISM: supernova remnants}

\section{Introduction}

Core-collapse supernovae which show minimal hydrogen are classified as Type IIb supernovae (SNe~IIb). The progenitor systems of these supernovae are currently not well known but have been modeled as either explosions of interacting binary systems with initial masses between $10$--$15\unit{M_\sun}$ \citep[e.g.,][]{Yoon2017,Eldridge2018} or single stars with initial masses above $20\unit{M_\sun}$ that have experienced Wolf-Rayet driven mass-loss immediately prior to explosion \citep[e.g.,][]{ChevalierSoderberg2010}. From pre-explosion images some SN~IIb progenitors have been identified as yellow supergiants (e.g., \citealp[SN~1993J, ][]{Aldering1994}; \citealp[SN~2011dh, ][]{Maund2011,VanDyk2011}; \citealp[SN~2013df,  ][]{VanDyk2014}; and \citealp[SN~2016gkg, ][]{Tartaglia2017}). In addition, for a handful of systems surviving companion candidates have been identified (e.g., \citealp[SN~1993J, ][]{Maund2004,Fox2014}; \citealp[SN~2001ig, ][]{Ryder2006,Ryder2018}; and \citealp[SN~2011dh, ][] {Folatelli2014,Maund2015}).

Investigations into a young supernova remnant's surrounding circumstellar material (CSM) can provide valuable information about the progenitor's pre-supernova evolution including its mass-loss history. Furthermore, the structure of young Galactic supernova remnants (SNRs) can be studied at much higher spatial resolutions than SNe at extragalactic distances.

With a current age of about $350\unit{yr}$ \citep{Thor01} and an estimated distance of $\sim3.4\unit{kpc}$ \citep{Minkowski59,Reed95,Alarie14}, Cassiopeia~A provides perhaps the clearest view of a high-mass progenitor, core-collapse supernova remnant. The Cas~A supernova appears to have been a Type IIb based on light echo spectra which showed strong spectral similarities to the prototypical Type IIb event SN~1993J \citep{Krause08,Rest08,Rest11}. No surviving companion has been identified for Cas~A, suggesting that if the system was a binary then either the two stars merged before the explosion, or the companion is a relatively faint, low mass dwarf star \citep{Kochanek18,Kerzendorf2019}.

The Cas~A remnant's main structure consists of an X-ray, optical, infrared, and radio bright emission shell of metal-rich ejecta with an angular radius of approximately $2\farcm 5$, corresponding to a linear diameter of $5\unit{pc}$ at $3.4\unit{kpc}$. The remnant's reverse shock-heated ejecta exhibit radial velocities between $-4500$ and $+6000 \unit{km}\unit{s}^{-1}$ \citep[e.g.,][]{Minkowski59,Reed95,Delaney10,MF13}. 

The remnant also possesses much lower velocity optical emission clumps known as `Quasi-Stationary Flocculi' (QSFs) thought to be shocked clouds of pre-supernova mass-loss circumstellar material. More than 100 QSFs have been identified, with the majority exhibiting negative radial velocities, while the full observed velocity range is ${+}100$ to $-550\unit{km}\unit{s}^{-1}$ \citep{vandenBerghKamper85,Alarie14,Koo2018}. These clumps are relatively dense ($\rm{n_e}=10^{3-4}\unit{cm}^{-3}$), exhibit unusually strong lines of nitrogen relative to hydrogen, and  based on proper motion studies are thought to be mass-loss material from the progenitor ${\sim}10^{4}$ years before outburst \citep[e.g.,][]{PeimbertvandenBergh71,KampervandenBergh76,CK78,vandenBerghKamper85}. 

The overabundance of nitrogen in the QSFs and the subsequent discovery of high-velocity nitrogen-rich but hydrogen-poor ejecta knots led \citet{FBB87} and \citet{FB91} to conclude that the progenitor's hydrogen and nitrogen-rich envelope was not completely removed prior to the supernova event. This meant that the Cas~A supernova could have exhibited H$\alpha$ emission at time of the outburst, a notion subsequently confirmed by light echo studies \citep{Krause08,Rest08,Rest11}.

Farther outside the remnant's high-velocity ejecta shell and QSF clumps lie faint emission clouds first noted by Minkowski in the 1960s. This emission, primarily seen northwest and northeast of Cas~A within 7--10\arcmin\ of the remnant's center \citep[see][]{vandenBergh71a}, was initially interpreted as part of a local, low density H~II region which was photoionized by the UV and X-ray radiation from the Cas~A supernova due to the lack of photoionizing OB stars in the remnant's local vicinity \citep{vandenBergh71a,Peimbert71}. 

However, \citet{CK78} and \citet{CO03} suggest these surrounding nebulae might be mass-loss material which pre-dated the mass-loss episode which formed the QSFs. Optical spectra taken of one of the eastern emission clouds showed surprisingly strong [\ion{S}{2}] line emission like that typically seen in shocked gas, despite being outside the remnant's forward shock \citep{FBB87}. It also exhibited weaker [\ion{N}{2}] line emissions than the QSFs, consistent with the picture of an earlier, less nitrogen-rich mass-loss episode.

Here we present the results of a much deeper and more extensive optical study of the extended interstellar environment toward and around Cas~A. We find that the nebular emission surrounding the remnant to the north and east is likely progenitor mass-loss material pre-dating the formation of the dense circumstellar QSFs. This material appears to have accumulated to the north and east as it expanded and collided with the outer portions of a neighboring H~II region. Our observations and results are presented in \S\S\ref{sec:Obs} and \ref{sec:Results}, respectively, with the implications regarding the supernova remnant and progenitor's mass-loss history discussed in \S\ref{Sec:Disc}. Our conclusions are summarized in \S\ref{sec:Con}.

\section{Observations}
\label{sec:Obs}

\begin{deluxetable*}{lcccccc}
\tablecaption{Spectroscopic Observations \label{table:specobs}}
\tablehead{
\colhead{Region Name} & \colhead{Date} & \colhead{Telescope} & \colhead{Instrument} & \colhead{Exposure} &\colhead{Wavelength} \vspace{-0.1cm} & \colhead{Resolution}\\ 
& & & & \colhead{Time (s) } & \colhead{Coverage (\AA)} & \colhead{(\AA)}}
\startdata
NE1 & 2004 Sept & 2.4m MDM & Modspec & 2 x 1000 & 5500 -- 8000 & 5.0 \\
NE2 & 2017 Oct & 2.4m MDM & OSMOS & 2 x 2000 & 4750 -- 6850 & 3.5\\
NE3 & 2018 Nov & 6.5m MMT & Binospec & 3 x 1000 & 4450 -- 9800 & 5.0\\
NW1 & 2004 Sept & 2.4m MDM & Modspec & 2 x 1200 & 5500 -- 8200 & 5.5\\
NW2 & 2004 Sept & 2.4m MDM & Modspec & 2 x 1500 & 5500 -- 7700 & 5.0\\
SE1 & 2017 Dec & 2.4m MDM& OSMOS &  1 x 1500 & 6200 -- 7250 & 8.0\\
SE2 & 2017 Dec & 2.4m MDM & OSMOS & 1 x 2000 & 6200 -- 7250 & 9.0\\
SW & 2004 Sept& 2.4m MDM & Modspec & 2 x 1200 & 5500 -- 7500 & 5.5\\
E1 -- E3 & 2018 Nov & 6.5m MMT & Binospec & 3 x 1000 & 4450 -- 9800 & 5.0\\
H~II Region & 2004 Sept & 2.4m MDM & Modspec & 2 x 1000 & 5500 -- 7650 & 5.0
\enddata
\end{deluxetable*}

\subsection{Imaging}
With the goal of placing the Cas~A supernova remnant in context with its interstellar environment, wide-field images were taken using the 0.8m McDonald Observatory telescope with the Prime Focus Corrector camera. This instrument used a Loral Fairchild $2048 \times 2048$ CCD, with a pixel size of $1\farcs355$, producing a field-of-view of $46\farcm2 \times 46\farcm2$. A series of images were taken in November 2003 using H$\alpha$ (6568\,\AA, FWHM=30\,\AA) and red continuum (6510\,\AA, FWHM=30\,\AA) filters at nine overlapping regions leading to a $1\fdg3\times1\fdg4$ mosaic image centered on the remnant. Exposure times ranged from $2$--$4 \times 1000\unit{s}$ to $2$--$3 \times 1200 \unit{s}$ for each region. 

\begin{figure*}[ht!]
   \centering
   \includegraphics[width=0.648 \textwidth]{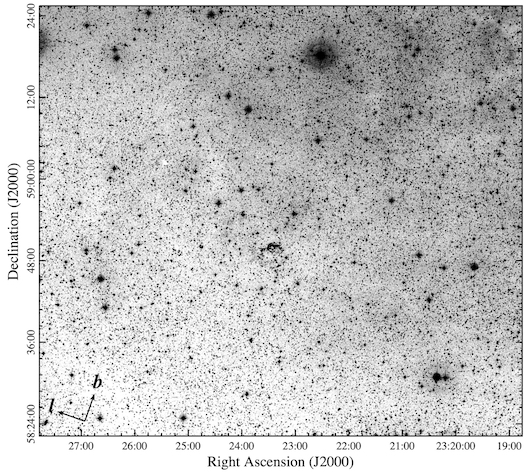} \\
   \includegraphics[width=0.648\textwidth]{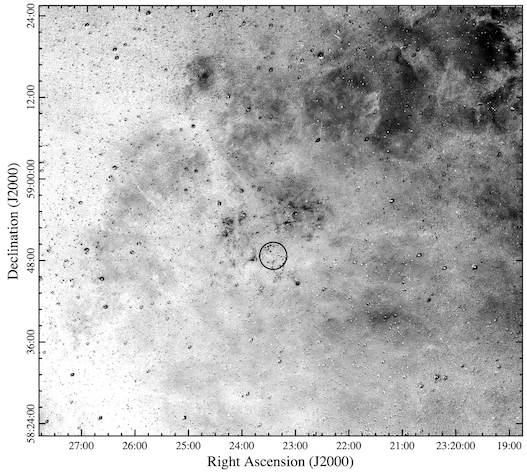}
   \caption{\textit{Top}: Broadband red DSS2 image of Cas~A  ({\it{l}} $= 111\fdg7$, {\it{b}} $= -2\fdg1$). The Galactic plane is toward the northwest as indicated by the compass in the lower left. \textit{Bottom}: H$\alpha$ minus red continuum mosaic image of this same region showing its projected interstellar environment. Cas~A's optical shell of high-velocity ejecta is represented by the black circle four arcminutes in diameter.  \label{fig:MOS_LARGE}}
\end{figure*}

\begin{figure*}[ht!]
   \centering
   \includegraphics[width=0.9\textwidth]{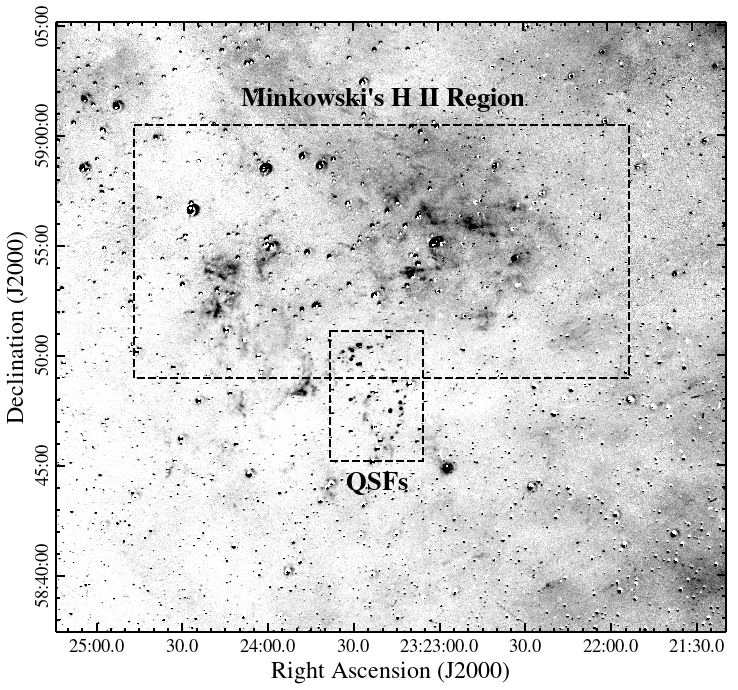}
   \caption{Enlargement of a section of the lower panel in Figure~\ref{fig:MOS_LARGE} showing the H$\alpha$ emission immediately around Cas~A. The faint diffuse emission first noted by Minkowski is marked by the large dashed black box, along with a region to the south centered on Cas~A and containing the QSFs. \label{fig:MinkowskiHII}}
\end{figure*}

\begin{figure*}[hbt!]
  \centering
   \includegraphics[width=0.32\textwidth]{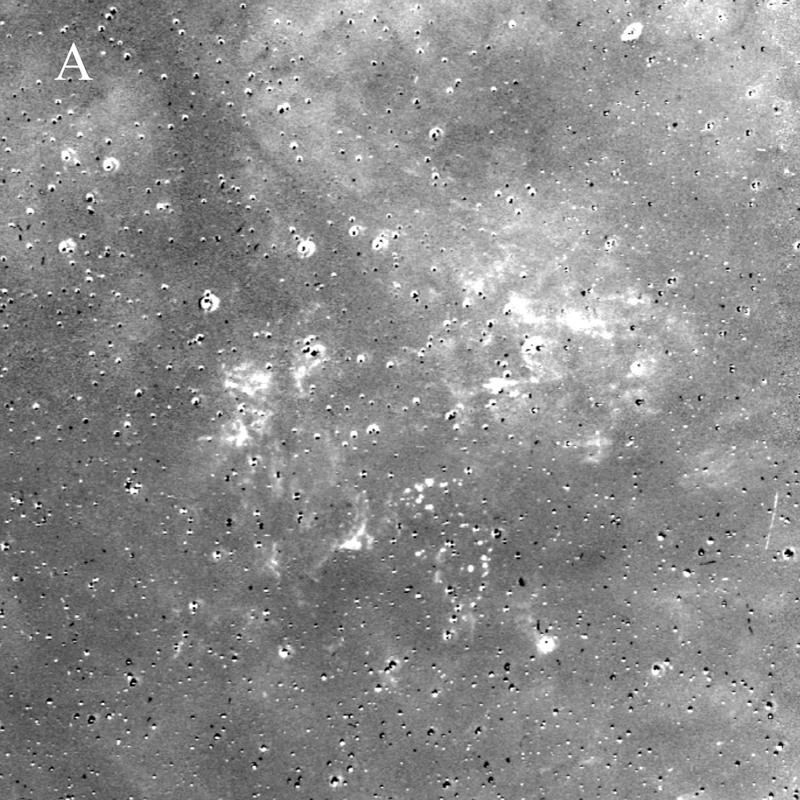}
   \includegraphics[width=0.32\textwidth]{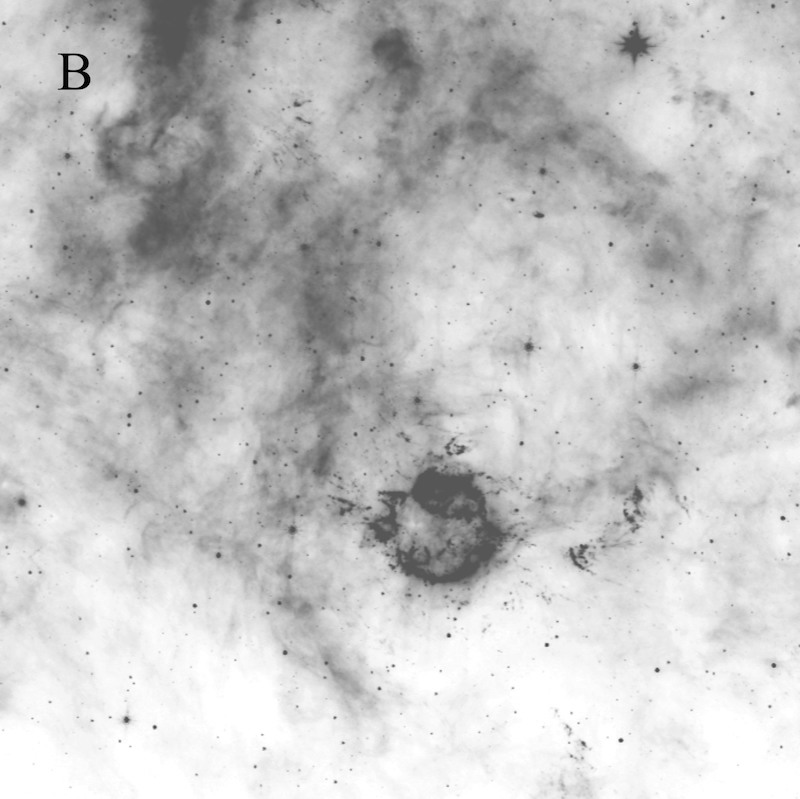}
   \includegraphics[width=0.32\textwidth]{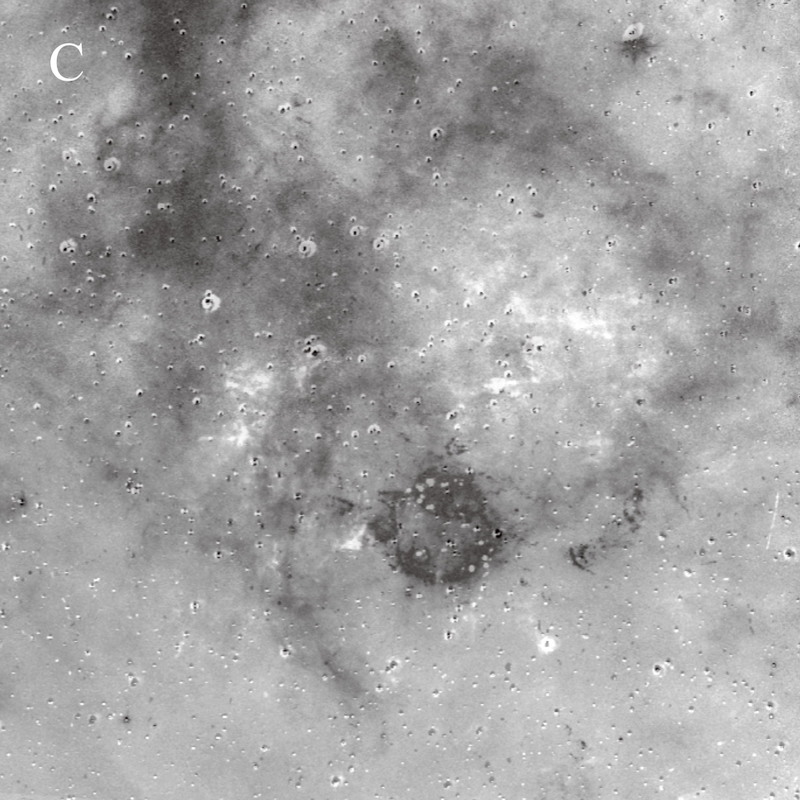}
   \caption{A positive H$\alpha$ minus continuum image (A, left), a negative {\sl Spitzer} IRAC $8\unit{\mu m}$ image taken 28 Dec 2006 (B, center; PI: Rudnick, ID: 30153), and a composite image of the H$\alpha$ minus continuum and {\sl Spitzer} images (C, right). The brightest H$\alpha$ emission lies coincident with weak {\sl Spitzer} emission.  \label{fig:CasAOptDust}}
\end{figure*}

Data reduction consisted of bias subtraction, flat-fielding, cross-registering, scaling, and continuum subtraction using the procedure described in \citet{GF07}. Continuum subtracted images were then combined using Montage\footnote{http://montage.ipac.caltech.edu} \citep{Montagecite}, to correct any residual background differences after subtraction from flat-fielding yielding a $\sim1\fdg0\times 1\fdg2$ region centered on Cas~A (see Fig.~\ref{fig:MOS_LARGE}). 

Follow-up, higher-resolution H$\alpha$ and red continuum images were taken in September 2017 at MDM Observatory to study regions which showed interesting morphology in the wide-field mosaic image. Images were taken using the Hiltner 2.4m telescope equipped with the Ohio State Multi-Object Spectrograph \citep[OSMOS;][]{Martini2011} and ITL $4064 \times 4064$ CCD. This instrument has an $18\farcm5 \times 18\farcm5$ field-of-view with a pixel size of $0\farcs273$. H$\alpha$ (FWHM=100\,\AA) observations were $3 \times 4000\unit{s}$ exposures, while continuum (6450\,\AA, FWHM=100\,\AA) observations consisted of $3$--$4 \times 2000\unit{s}$ exposures. 

Additional images were taken in October 2017 using the McGraw-Hill 1.3m telescope and a LBNL $4096 \times 4096$ CCD. This telescope-camera system provided a $21\farcm3 \times 21\farcm3$ field-of-view with $0\farcs 315$ pixels. An H$\alpha$ image was obtained by combining five overlapping regions from a total of $17 \times 1200\unit{s}$ exposures, while an off-band continuum image was made from five overlapping regions consisting of $7 \times 1200\unit{s}$ exposures. Because the 6450\,\AA\ continuum filter image had contamination from Cas~A's high-velocity [\ion{O}{1}] 6300\,\AA\ ejecta, this emission was digitally removed to match background levels in the H$\alpha$ minus continuum image. 

\subsection{Spectroscopy}
Low-dispersion spectra were taken of several H$\alpha$ emission features in September 2004 using the Modspec spectrograph on the MDM Observatory Hiltner 2.4m telescope with a $1\farcs1$ slit and the Echelle CCD, a SITe $2048 \times 2048$ CCD (see Table~\ref{table:specobs}). Observations were also taken in October and December 2017 using OSMOS and the ITL $4064 \times 4064$ CCD. Observations in October 2017 used a $1\farcs2$ slit with a blue VPH grism, while observations taken in 2017 December employed a $1\farcs4$ slit with a red VPH grism (see Table~\ref{table:specobs}). Spectra were extracted with sizes between 2--3\arcsec\ along the slit concentrating on the brightest emission regions. 

Standard pipeline data reduction was performed using PYRAF\footnote{PYRAF is a product of the Space Telescope Science Institute, which is operated by AURA for NASA.}. The spectra were bias-subtracted, cosmic-ray corrected using the L.A.\ Cosmic software \citep{vanDokkum01}, co-added, wavelength calibrated using Hg, Ne, Ar, and Xe comparison lamps, and flux calibrated using multiple standard stars. 

Additional spectra were obtained with the 6.5m MMT using Binospec \citep{Fabricant2019}, an imaging spectrograph, in November 2018. Data were obtained using a 1\arcsec\ slit and the 270 line grating (see Table~\ref{table:specobs}), with individual spectra extracted at four positions along the slit with sizes between 2--3\arcsec. These data were reduced using the Binospec pipeline for bias-subtraction, flat-fielding, cosmic-ray correction and wavelength calibration \citep{BinospecCite}. The spectra were flux calibrated with a standard star. Uneven continuum levels remaining after sky subtraction and flux calibration were removed by subtracting the underlying continuum fit. Residual night-sky line features were also removed from the final spectra.

\section{Results} 
\label{sec:Results}
\subsection{H$\alpha$ Imaging}
\label{sec:HAimage}
\begin{figure*}[h!]
   \centering
    \includegraphics[width=0.495\textwidth]{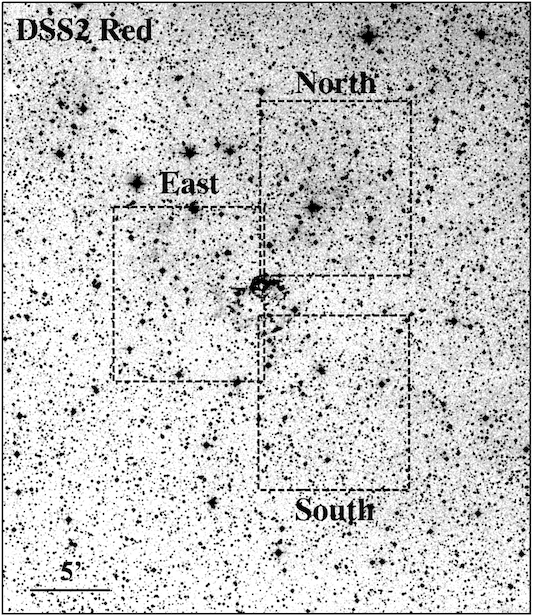} 
    \includegraphics[width=0.495\textwidth]{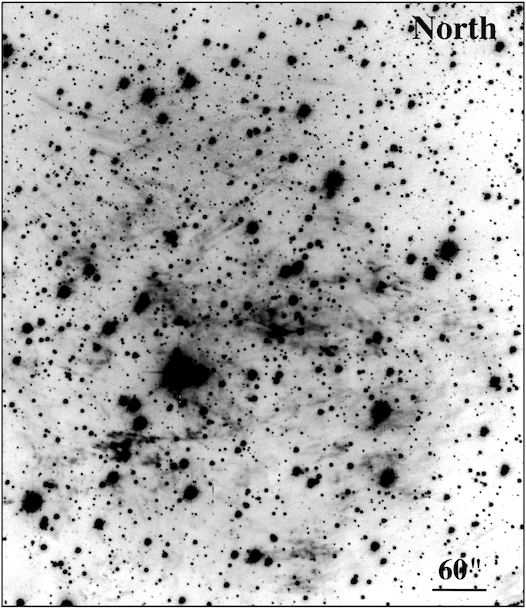}\\
    \includegraphics[width=0.495\textwidth]{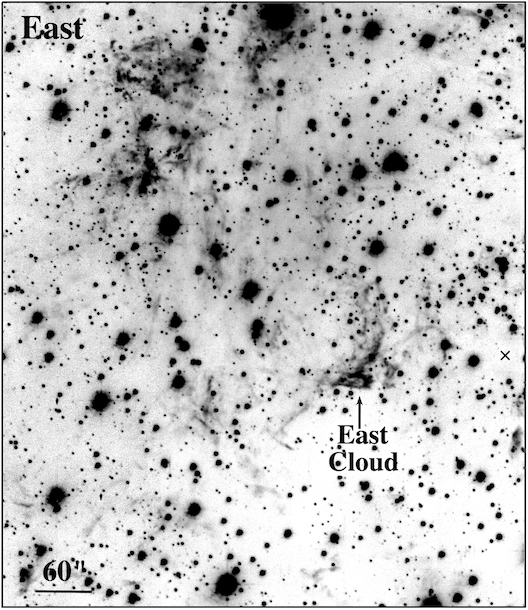}
    \includegraphics[width=0.495\textwidth]{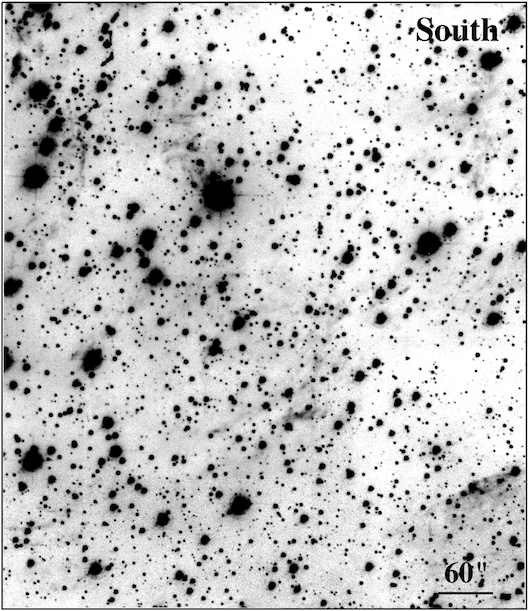}
    \caption{DSS2 red image (upper left) of Cas~A's high-velocity, metal-rich ejecta shell and the surrounding interstellar emission with dashed boxes showing regions where 2.4m Hiltner Telescope high resolution H$\alpha$ images were obtained. We will refer to the brightest emission feature in the east region (lower left) as the `East Cloud.' The center of expansion is marked by an ``x,'' on the the right side of the east region. North is to the top, East is to the left. \label{fig:HighRes}}
\end{figure*}

The top panel of Figure~\ref{fig:MOS_LARGE} shows a red DSS2 image of a $\sim1\fdg0\times 1\fdg0$ area around Cas~A. The remnant's $\sim4\arcmin$ diameter emission shell of reverse-shock heated, high-velocity metal-rich, hydrogen-poor ejecta is clearly visible in this image along with faint diffuse emission around the remnant. The bottom panel of Figure~\ref{fig:MOS_LARGE} shows this same area in H$\alpha$ emission. The remnant's hydrogen-poor, metal-rich emission shell is represented by the circle seen near the center of the image. Bright almost point-like emission features seen inside the circle are the remnant's circumstellar mass-loss QSFs. Note at Cas~A's distance of $3.4\unit{kpc}$, 1\arcmin\ is equal to $1\unit{pc}$.

\begin{figure*}[htb!]
   \centering
    \includegraphics[width=0.495\textwidth]{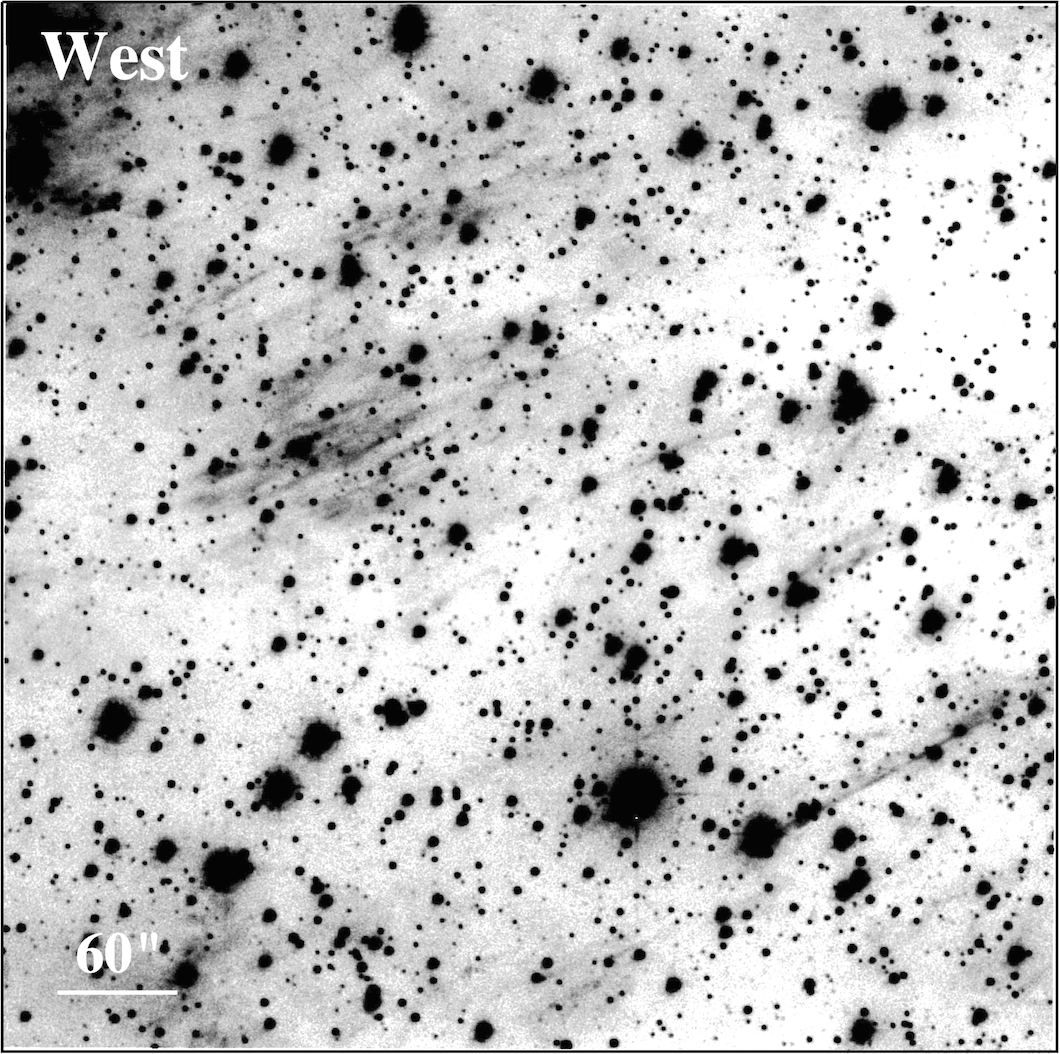}
    \includegraphics[width=0.495\textwidth]{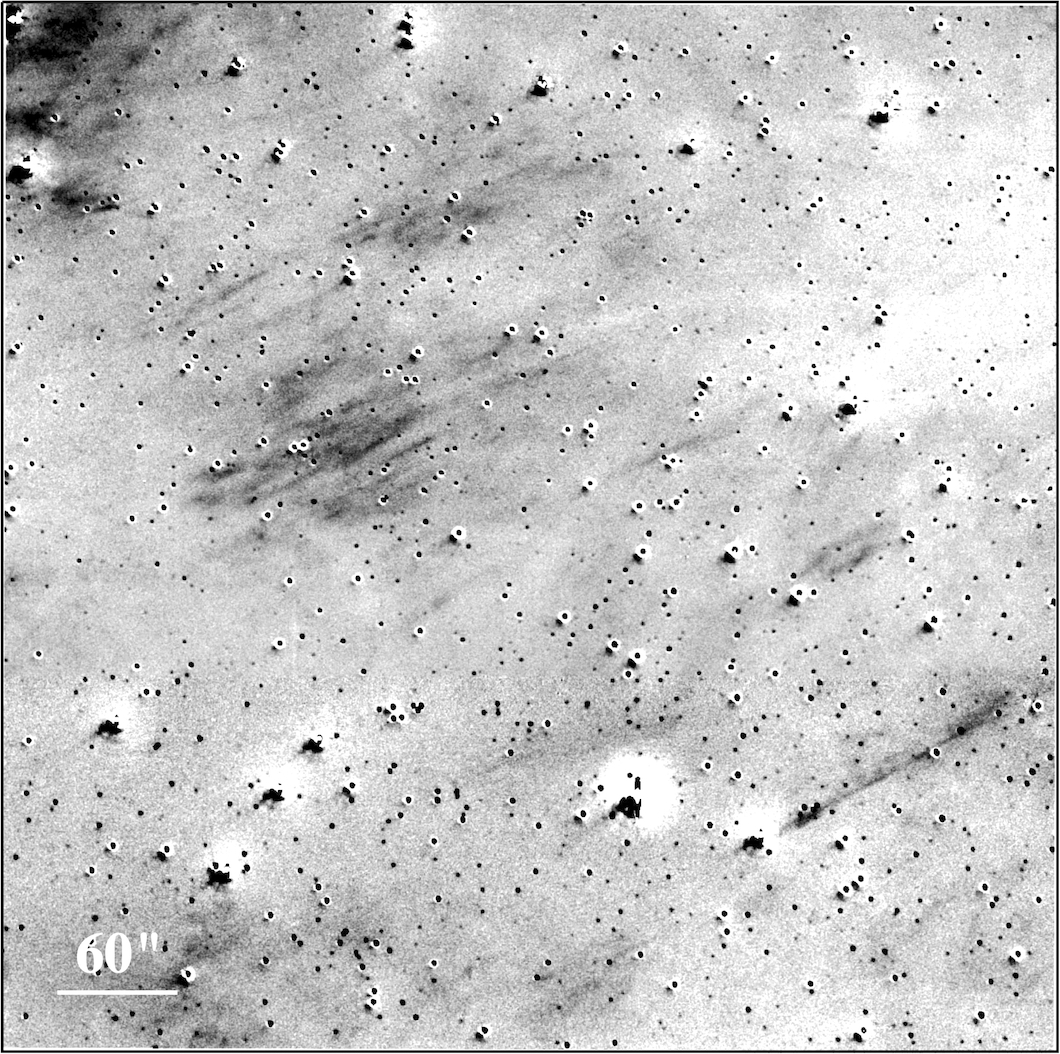} 
   \caption{H$\alpha$ (left) and H$\alpha$ minus continuum (right) images an $8\farcm8 \times 8\farcm8$ region west of the remnant showing parallel emission `streaks.' North is to the top, East is to the left. \label{fig:Streaks}}
\end{figure*}

\begin{figure*}[bht!]
   \centering
    \includegraphics[align=t,height=0.55\textheight]{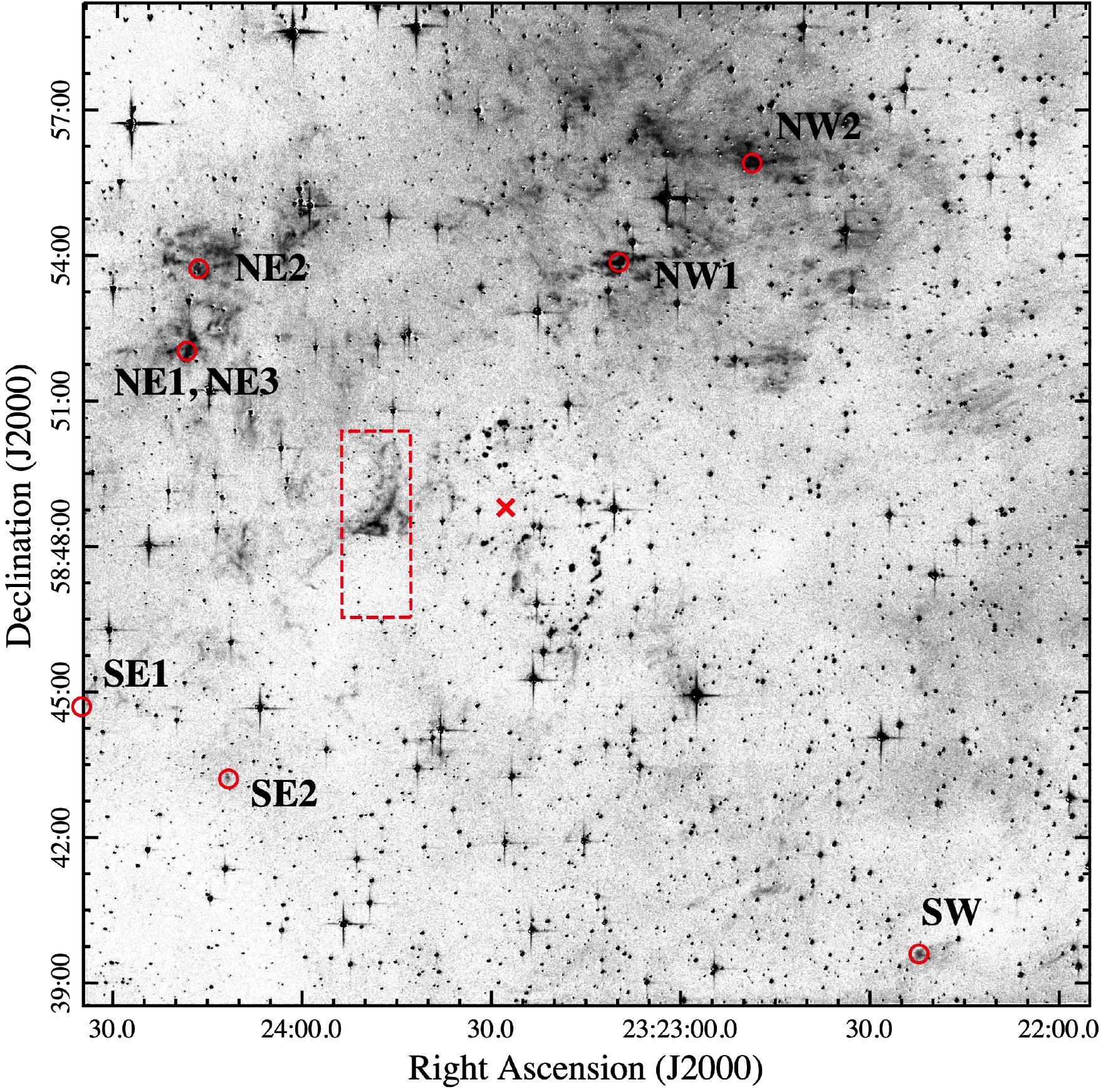}
    \includegraphics[align=t,height=0.507\textheight]{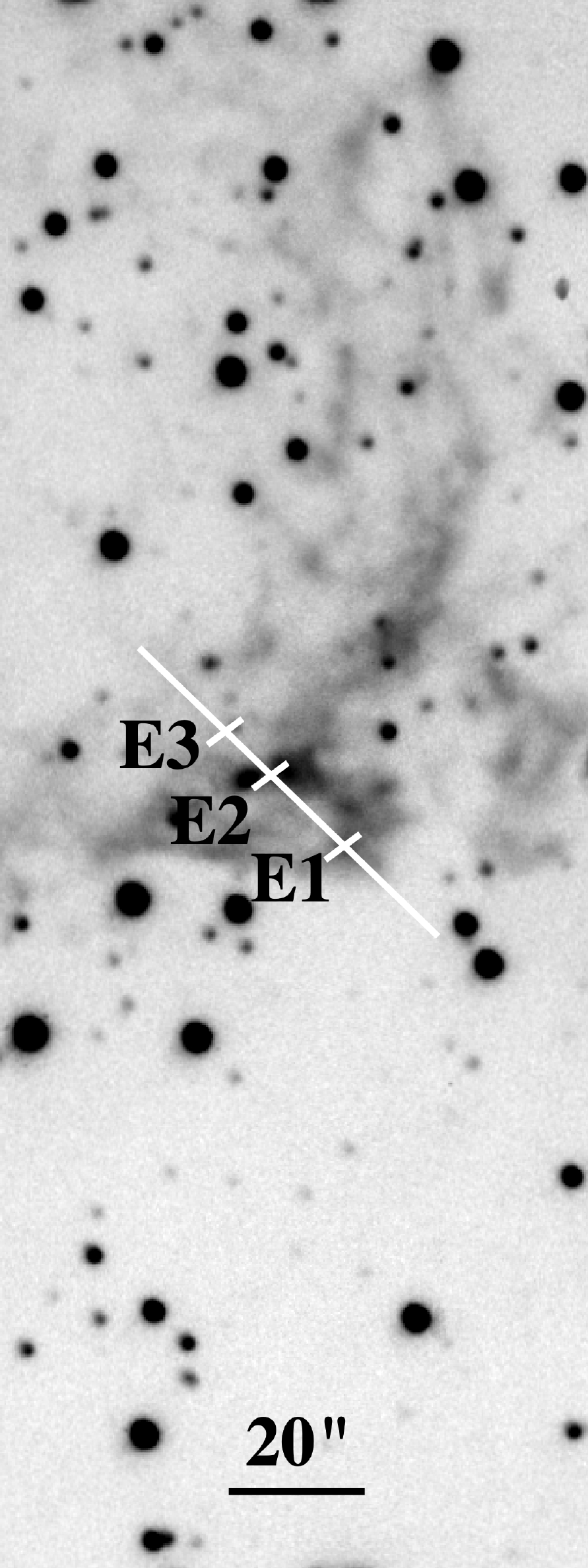}
   \caption{\textit{Left}: H$\alpha$ minus continuum image with slit positions marked by open red circles. The center of expansion is marked with an ``x.'' The continuum image used to remove stars has contamination from high-velocity [\ion{O}{1}] 6300, 6364\,\AA\ seen in the optical shell of Cas~A, which has been corrected to match the background level. \textit{Right}: H$\alpha$ image of the boxed region on the left, showing the East Cloud slit positions. \label{fig:specpos}}
\end{figure*}

On large scales, this image reveals the presence of considerable extended diffuse emission broken up in places by several dust lanes and clouds. This emission, brightest about 30\arcmin\ to the northwest, represents the outskirts of an H~II region centered approximately 1.5 degrees to the northwest. 

However, on smaller scales, clumpy emission about 10\arcmin\ to 15\arcmin\ in extent is seen northwest and northeast of Cas~A but notably absent to the remnant's south and southwest. This emission is brighter, clumpier and hence morphologically distinct from the more extended and more diffuse neighboring H~II region emission. 

These close in emission features can be better seen in Figure~\ref{fig:MinkowskiHII} and correspond to the perceived faint H~II region reported by Minkowski \citep[hereafter `Minkowski's H~II region'; see][]{vandenBergh71a}. In this image, the nebulosities can be seen as clumps and filaments with sizes on the order of 1--2\arcmin, along with fainter and more diffuse surrounding emission. In sharp contrast, few if any similar emission features are present south of the remnant. Also marked on Figure~\ref{fig:MinkowskiHII} is the location of the remnant's brighter QSFs. Since each QSF is only a few arcseconds in size, they appear as small, semi-stellar features arranged in a distinctly non-spherical structure.

We have found evidence that the clumpy emission north and east of Cas~A is significantly affected by interstellar medium (ISM) dust along our line-of-sight. This is shown in the series of images presented in Figure~\ref{fig:CasAOptDust} which shows a continuum subtracted H$\alpha$ image (left) as seen in Figures~\ref{fig:MOS_LARGE} and \ref{fig:MinkowskiHII}, a negative {\sl Spitzer} IRAC $8\unit{\mu m}$ image (center) of the same region, and an overlay of the H$\alpha$ minus continuum and $8\unit{\mu m}$ images (right). 

Comparison of these images reveals that apparent breaks or holes in the clumpy nebulosity surrounding the remnant are, in fact, largely coincident with the $8\unit{\mu m}$ emission, which traces the ISM dust along the line-of-sight. This means that there is likely a more continuous H$\alpha$ emission structure north and east of Cas~A than visible in Figures~\ref{fig:MOS_LARGE} and \ref{fig:MinkowskiHII}. 

Not believed part of these clumpy nebulae is a chain of small, bright emission clumps immediately west of Cas~A seen only in this epoch of the {\sl Spitzer} $8\unit{\mu m}$ image (middle panel of Fig.~\ref{fig:CasAOptDust}). These features could be a light-echo from the supernova explosion \citep{Krause08,Rest08,Rest11}. In this case, the dust producing this light echo would have to be a large distances from the remnant and thus is unlikely to be associated with mass-loss from the progenitor.  

Higher-resolution H$\alpha$ images of regions around the remnant are shown in Figures~\ref{fig:HighRes} \& \ref{fig:Streaks}. Prior to these images, the deepest image of the interstellar environment around Cas~A was that presented in Figure~10 of \citet{Fesen01}. That image's small field-of-view only detected the remnant's QSFs and a small emission cloud east of the remnant, thus missing the extended emission clouds farther to the north and northeast of the remnant. 

The upper left panel of Figure~\ref{fig:HighRes} shows the broadband red DSS2 image of Cas~A and its surroundings. Visible in this image is Cas~A's optical shell of metal-rich, high-velocity ejecta, along with much fainter diffuse circumstellar emission to the north and east.

The upper right panel of Figure~\ref{fig:HighRes} shows the region immediately north and northwest of Cas~A containing some of the largest and brightest emission clouds adjacent to the remnant. In contrast, much fainter, scattered and more diffuse emission is seen south and southwest of the remnant in the lower right panel of Figure~\ref{fig:HighRes}. The southern extension of the QSFs is visible in the upper-left of this panel, along with a faint north-south emission filament extending approximately 45\arcsec\ off the remnant's southwest limb. Previous H$\alpha$ imaging of Cas~A only detected the northern most part of this north-south emission filament, and barely detected the cloud to the west \citep[see Fig.~10 of][]{Fesen01}. 

The closest H$\alpha$ emission cloud to Cas~A's main shell of ejecta is a triangular shaped emission patch located about one arcminute due east (see lower left panel of Fig.~\ref{fig:HighRes}). Previously referred to by some as the eastern nebulosity, we will instead refer to this emission feature as the `East Cloud' to distinguish it from the other diffuse emission in the east.

This East Cloud is located just outside the remnant's forward shock front as measured by the remnant's outer nonthermal and X-ray emitting filaments \citep{Gotthelf01,DeLaney03,Patnaude09,Delaney10}. In this area, Cas~A's highest velocity ejecta knots move ballistically and have outrun the forward shock and thus appear bright due to their direct interaction with this cloud (see \S\ref{sec:DiscMink}). In fact, previous imaging by \citet{FBB87} showed that parts of the East Cloud extend toward the north, coincident with the region of Cas~A's northeast jet which may play a role in the visibility of the jet's emission knots.  

Several arcminutes farther to the northeast one finds additional emission patches of similar size and brightness (see lower left panel of Fig.~\ref{fig:HighRes}) along with considerable diffuse emission. Although these emission clouds are visible in Figures~\ref{fig:MOS_LARGE} and \ref{fig:MinkowskiHII}, this image reveals much finer details including the extent and structure of this emission complex. These northeast H$\alpha$ clouds appear morphologically similar to both the East Cloud and the emission clouds north and northwest of Cas~A. 

A different and quite unexpected emission morphology was found west of Cas~A. A series of roughly parallel emission filaments or `streaks' are seen to the west-northwest of the remnant (see the left panel of Fig.~\ref{fig:Streaks}). The image is centered approximately 12 arcminutes west of the expansion center. The right panel of the figure shows the same region after continuum subtraction, where the continuous nature of these parallel emission features is more clearly visible. At a distance of $3.4\unit{kpc}$, like that of Cas~A, individual streaks would be approximately $0.5$ to $2\unit{pc}$ in length. Whereas the more northern streaks point in the general direction of Cas~A's expansion center, the southern ones do not.

\begin{figure*}[ht!]
  \centering
   \includegraphics[width=0.7\textwidth]{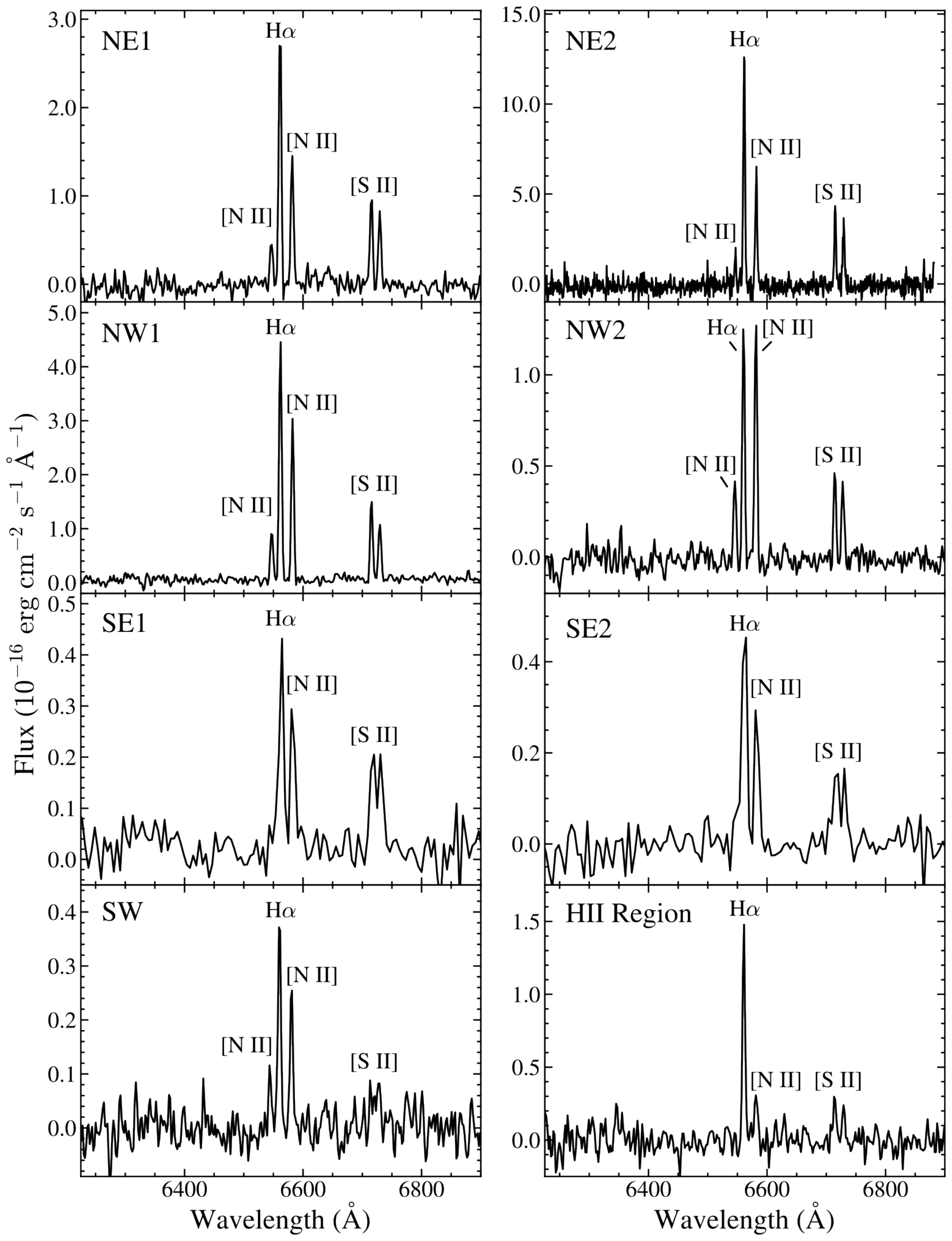}
   \caption{Eight CSM spectra from regions around Cas~A, shown in Figure~\ref{fig:specpos}. The `H~II Region' spectrum is from the diffuse H~II region, $\sim$50\arcmin\ northwest of the center of expansion (see Fig.~\ref{fig:MOS_LARGE}). \label{fig:Spectra}}
\end{figure*}

\begin{figure}[th!]
   \centering
    \includegraphics[width=\columnwidth]{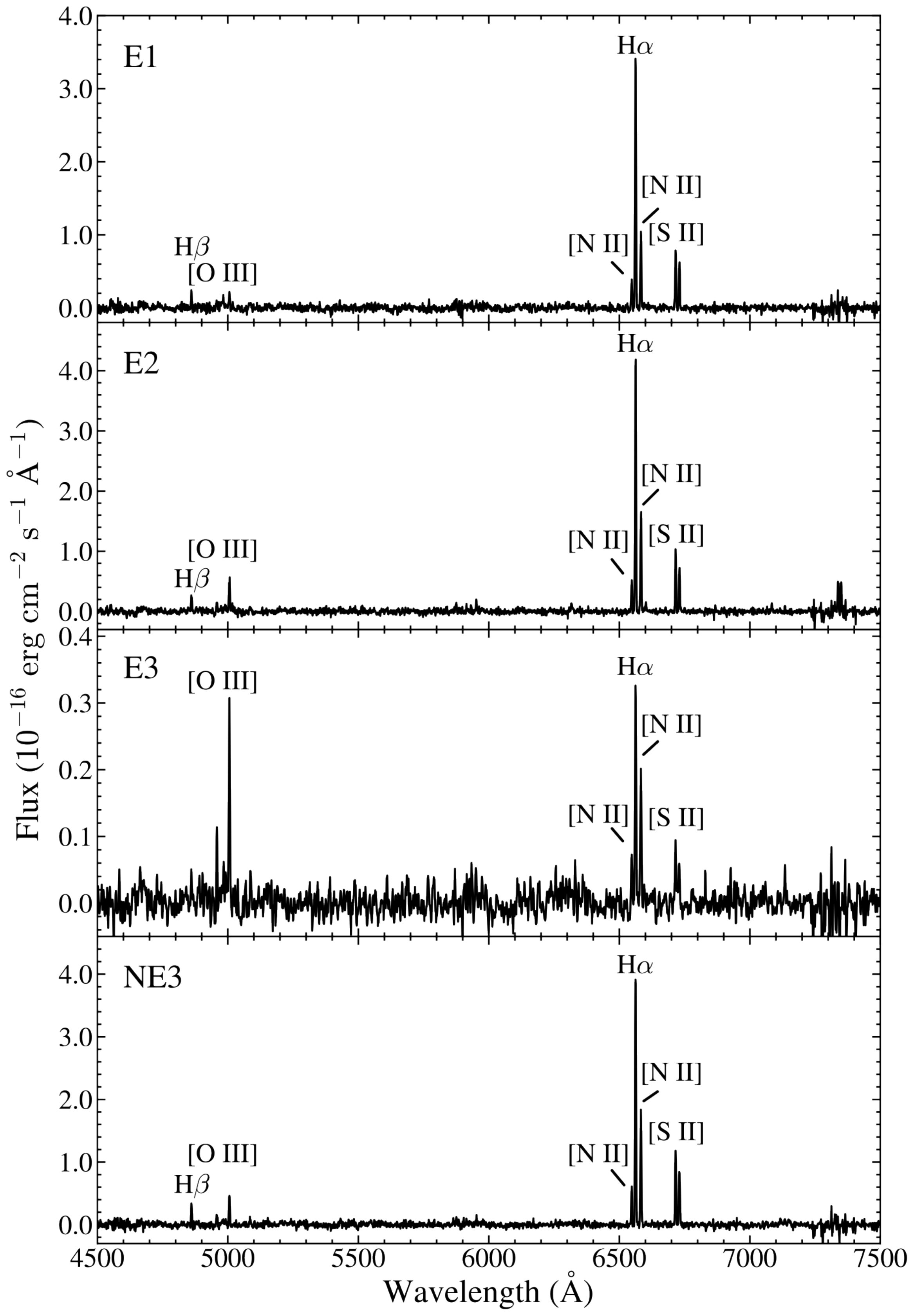}
   \caption{Diffuse nebulosity spectra of the north-east region of Cas~A. Residual night sky-lines at 5577\,\AA\ and 6300\,\AA, have been removed after imperfect sky subtraction. Note: E2 has an overlapping high-velocity oxygen and nitrogen rich ejecta knot. \label{fig:mmtspec}}
\end{figure}

\subsection{Optical Spectra}

Low-dispersion optical spectra were obtained for a dozen emission features around the remnant. Exact slit positions are shown in Figure~\ref{fig:specpos}, with the resulting spectra presented in Figures~\ref{fig:Spectra} and \ref{fig:mmtspec}. Relative emission line fluxes for the 12 spectral locations are presented in Table~\ref{table:linemeas}. The H~II region spectrum shown sampled a region located roughly 50\arcmin\ northwest of the remnant (see Fig.~\ref{fig:MOS_LARGE}). Except for the H~II region, spectra of all emission regions surrounding Cas~A exhibit similar line emission intensity ratios. Moreover, spectra of features located north of Cas~A, the East Cloud closest to the remnant, and filaments in the southeast and southwest differ substantially from that of QSFs. 

Whereas QSFs show [\ion{N}{2}] 6583\,\AA\ to H$\alpha$ ratios $\approx3$, indicating an overabundance of nitrogen by an order of magnitude, and densities of approximately $10^{3-4}\unit{cm}^{-3}$ \citep{KC77,CK78,HF96,Alarie14}; spectra of emission clouds surrounding Cas~A show much weaker [\ion{N}{2}] line emissions relative to H$\alpha$, with [\ion{S}{2}] 6716/6731 line ratios indicating densities below $500\unit{cm}^{-3}$.

Surprisingly, all show [\ion{S}{2}]/H$\alpha\geq 0.4$ like that commonly seen in evolved SNRs due to shocked ISM. This is despite the fact that all lie outside Cas~A's forward shock front \citep{Delaney10}. However, none show appreciable [\ion{O}{1}] 6300, 6364\, \AA\ emission, which is a secondary indicator to discriminate photoionized H~II regions from shocked material like that present in SNRs \citep{Fesen85,Kopsacheili2020}.

Not unexpectedly, the brightest emission regions, have higher densities compared to the fainter emission regions which approach the low density limit for the [\ion{S}{2}] 6716/6731 ratio. As the distance away from the remnant increases, the ratio of [\ion{N}{2}] 6548, 6583\, \AA\ to H$\alpha$ decreases.

Our spectra of the East Cloud (E1, E2, \& E3) are in good agreement with a previously published spectrum of this cloud by \citet{FBB87} who found strong [\ion{N}{2}] emission relative to H$\alpha$ and a [\ion{S}{2}]/H$\alpha$ ratio of 0.5. Our spectra also show [\ion{S}{2}]/H$\alpha$ ratios between 0.4 and 0.57 with the highest value furthest from the remnant (E3) where [\ion{O}{3}] emission was also strongest. 

We estimated the extinction to the east of Cas~A assuming a Balmer ratio of H$\alpha$/H$\beta=2.9$, a temperature of $10^{4}\unit{K}$, and ignoring collisional contributions to the H$\alpha$ flux and hence our values should be viewed as upper limits. Assuming an R value of 3.1, we find extinction values for E1, E2, and NE3 to be: $E(B-V)=1.86$ $(A_{\rm V}\approx5.78\unit{mag})$, $E(B-V)=1.73$ $ (A_{\rm V}\approx5.34\unit{mag})$, and $E(B-V)=1.34$ $(A_{\rm V}\approx4.15\unit{mag})$, respectively. These values are in good agreement with extinction values found by \citet{HF96} for the northeast region around Cas~A, and imply that the [\ion{O}{3}]/H$\alpha$ ratios in Table~\ref{table:linemeas} should be multiplied by about a factor of 4, while the [\ion{S}{3}]/H$\alpha$ ratios should be divided by about 6.

\bigskip
\bigskip
\section{Discussion}
\label{Sec:Disc}
The interstellar environment around Cas~A can be used to trace the progenitor's mass-loss history and evolution right up to the time of the explosion. In the late 1960's, Minkowski found diffuse emission surrounding Cas~A and considered it be an H II region, possibly related to the progenitor's birth \citep{vandenBergh71a}. However, \citet{CK78} instead suggested that this emission might be pre-supernova mass-loss material. 

The ionization source for the faint emission north and east of the remnant was initially thought to be photoionization from OB stars in the local vicinity of Cas~A. However, the lack of such stars within 300\arcsec\ of the remnant led \citet{vandenBergh71a} and \citet{Peimbert71} to consider that the gas was instead ionized by X-rays from the supernova outburst. Below we discuss possible ionization sources and present evidence that these emission clouds are physically at Cas~A's distance and thus likely RSG mass-loss material.  

\subsection{Evidence for a Direct Association Between Neighboring Nebulae and Cas~A}
\label{sec:DiscMink}

The close proximity and spectral properties of the triangular-shaped East Cloud emission feature led \citet{FBB87} to suggest that it might be pre-supernova mass-loss material interacting with Cas~A's outer, high-velocity ejecta. Similarly, \citet{CK78}, \citet{CO03}, and \citet{chevalier05} advocated that this cloud, as well as clouds north of the remnant, were RSG mass-loss material from Cas~A's progenitor, ionized by emission during shock breakout.

However, a physical connection between these clouds and Cas~A cannot be established solely on projected proximity of emission features. None of the emission nebulae seen around Cas~A have known distances, thereby making associations between the projected nebulae and the Cas~A remnant uncertain. Here we present strong evidence the East Cloud is indeed located at the remnant's distance and is physically interacting with some of Cas~A's high-velocity ejecta.

It has been shown that when Cas~A's high-velocity ejecta knots pass through local or circumstellar gas, they can undergo significant brightening due to internal shocks  \citep{Fesen11}. This has been observed for a number of both high-velocity, metal-rich ejecta knots in Cas~A's northeastern jet as well as high-velocity, nitrogen-rich ejecta knots seen around much of the remnant's outer periphery.

\begin{figure}[t!]
  \centering
   \includegraphics[height=0.2825\textheight]{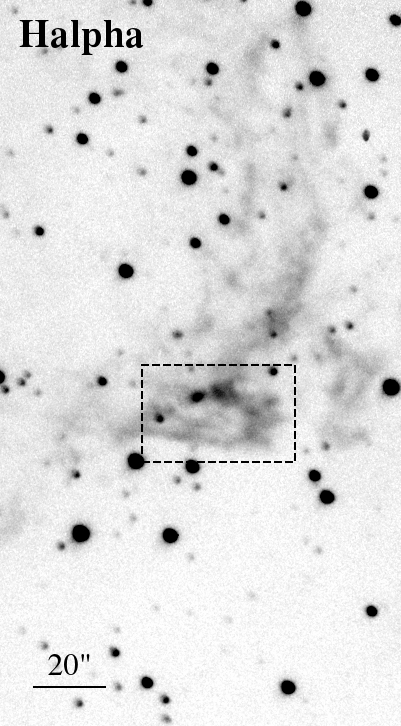}
   \includegraphics[height=0.2825\textheight]{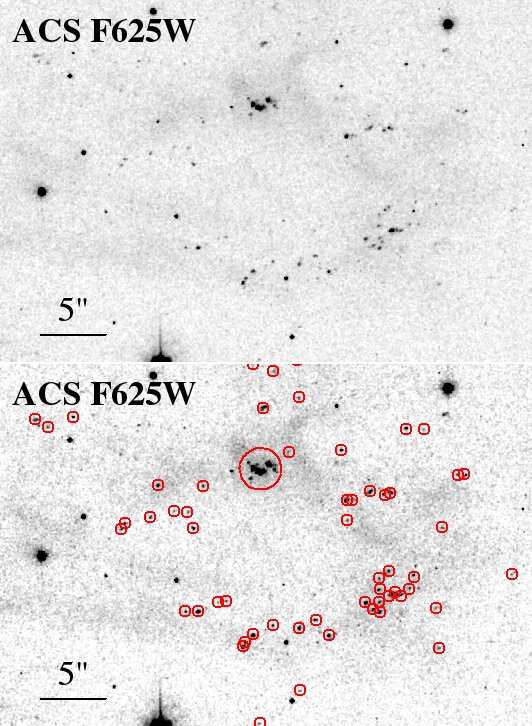}
   \caption{\textit{Left}: H$\alpha$ image of the East Cloud. \textit{Right}: {\sl HST} red image where ejecta knots are seen to brighten as they cross the East Cloud. Ejecta knots are marked with red circles. (PI: Fesen, GO: 9890, 15515). \label{fig:disEast}}
\end{figure}

This phenomena has been seen in {\sl HST} images taken of the East Cloud. Figure~\ref{fig:disEast} shows a {\sl HST} ACS F625W image of the East Cloud with the locations of coincident ejecta knots marked by red circles, identified through their proper motions. As can be seen in this figure, the majority of ejecta knots lie within the brightest emission regions of the East Cloud. Because the locus of these brightened ejecta knots matches the East Cloud's emission structure, the East Cloud must lie physically adjacent to Cas~A. Other surrounding H$\alpha$ emission clouds exhibit similar spectra and morphologies to that of the East Cloud, suggesting that they arise from a similar origin. Therefore, it is quite likely that they too lie at the same distance as Cas~A. 

There is also additional evidence to support this conclusion. Firstly, the right panel of Figure~\ref{fig:CasAOptDust} shows, a composite image with the {\sl Spitzer} IRAC 8$\unit{\mu m}$ emission and the H$\alpha$ minus continuum emission. The {\sl Spitzer} image traces the interstellar dust emission along the line-of-sight to Cas~A, and one notes that the brightest H$\alpha$ emission features lie in regions with relatively low or no 8$\unit{\mu m}$ emission. Further, the strong IR emission lies in between the bright H$\alpha$ emission features separating them into distinct emission regions. If there was no line-of-sight dust emission, it is likely that the emission to the north would appear as one continuous emission region. 

Secondly, a compact H~II region, IRAS 23151+5912, located approximately one degree northwest of the center of expansion, lies at an estimated distance of $3.33^{+1.23}_{-0.7}\unit{kpc}$ \citep{Choi14}. Assuming that this compact H~II region is part of the larger diffuse H~II region to the northwest of Cas~A (see Fig.~\ref{fig:MOS_LARGE}), then the extended H~II region lies at roughly the same distance as Cas~A. In this view, Cas~A's circumstellar emission is primarily seen north of the remnant because the progenitor's mass-loss contacted and gradually built-up as it ran into the outskirts of this H~II region.

\begin{deluxetable*}{lCCCCCCCCCCCC}
\tablecaption{Observed Relative Fluxes for Emission Features in Cas~A \label{table:linemeas}}
\tablehead{
\colhead{Line/Ratio} \vspace{-0.1cm}  & \colhead{NE1} & \colhead{NE2} & \colhead{NE3} & \colhead{NW1} & \colhead{NW2} & \colhead{SE1} & \colhead{SE2} & \colhead{SW} & \colhead{E1} & \colhead{E2} & \colhead{E3} &  \colhead{H II}\\
& & & & & & & & & & & & \colhead{Region}}
\startdata
H$\beta$ $\lambda$4861 & \nodata & \nodata & 8 & \nodata & \nodata & \nodata & \nodata & \nodata & 6 & 7 & \nodata & \nodata\\
$[$\ion{O}{3}$]$ $\lambda$4959 & \nodata & \nodata & 4 & \nodata & \nodata & \nodata & \nodata & \nodata & \nodata & 4 & 28 & \nodata\\
$[$\ion{O}{3}$]$ $\lambda$5007 & \nodata & \nodata & 12 & \nodata & \nodata & \nodata & \nodata & \nodata & 6 & 12 & 94 & \nodata\\
$[$\ion{N}{2}$]$ $\lambda$6548 & 9 & 11 & 14 & 23 & 39 & \nodata & \nodata & 23 & 12 & 13 & 24 & \nodata\\
H$\alpha$ $\lambda$6563 & 100 & 100 & 100 & 100 & 100 & 100 & 100 & 100 & 100 & 100 & 100 & 100\\
$[$\ion{N}{2}$]$ $\lambda$6583 & 53 & 53 & 48 & 64 & 104 & 64 & 75 & 67 & 30 & 38 & 66 & 27 \\
$[$\ion{S}{2}$]$ $\lambda$6716 & 31 & 34 & 28 & 33 & 39  & 57 & 47 & (28) & 24 & 23 & 33 & 25\\
$[$\ion{S}{2}$]$ $\lambda$6731 & 29 & 28 & 23 & 24 & 35  & 40 & 31 & (27) & 20 & 17 & 25 & 18\\
$[$\ion{Ar}{3}$]$ $\lambda$7136 & \nodata & \nodata & \nodata & \nodata & \nodata & \nodata & \nodata & \nodata & \nodata & 2 & 19 & \nodata\\
$[$\ion{O}{2}$]$ $\lambda$7320 & \nodata & \nodata & \nodata & \nodata & \nodata & \nodata & \nodata & \nodata & \nodata & \leq 1 & \nodata & \nodata\\
$[$\ion{S}{3}$]$ $\lambda$9069 & \nodata & \nodata & 11 & \nodata & \nodata & \nodata & \nodata & \nodata & 7 & 8 & 69 & \nodata\\
$[$\ion{S}{3}$]$ $\lambda$9531 & \nodata & \nodata & 20 & \nodata & \nodata & \nodata & \nodata & \nodata & 17 & 24 & 150 & \nodata\\
$[$\ion{S}{2}$]$/H$\alpha$ & 0.60 & 0.62 & 0.52 & 0.57 & 0.74 & 0.97 & 0.78 & 0.55 & 0.44 & 0.40 & 0.57 & 0.43\\
$\lambda$6716/$\lambda$6731 & 1.07 & 1.23 & 1.22 & 1.34 & 1.11 & 1.41 & 1.55 & \nodata & 1.22 & 1.31 & 1.33 & 1.40\\
$\rho$ (cm${}^{-3}$)\tablenotemark{a} & 450 & 200 & 200 & \leq 100 & 380 & \leq 100 & \leq 100 & \nodata & 200 & \leq 100 & \leq 100 & \leq 100\\ 
H$\alpha$ flux\tablenotemark{b} & 15.5 & 47.0 & 19.6 & 24.4 & 6.7 & 4.2 & 3.6 & 2.3 & 16.7 & 20.3 & 1.7 & 8.1\\
E(B--V)\tablenotemark{c} & \nodata & \nodata & 1.34 & \nodata & \nodata & \nodata & \nodata & \nodata & 1.86 & 1.73 & \nodata & \nodata
\enddata
\tablenotetext{a}{Electron densities are derived from the [\ion{S}{2}] $\lambda6716/\lambda6731$ ratio assuming $\rm{T}=10^{4}\unit{K}$.}
\tablenotetext{b}{Flux units: $10^{-16}\unit{erg}\unit{cm}^{-2}\unit{s}^{-1}$.}
\tablenotetext{c}{E(B--V) derived assuming H$\alpha$/H$\beta$=2.9, $\rm{T}=10^{4}\unit{K}$.}
\end{deluxetable*}

\subsection{Photoionization of Surrounding Nebulae}
\label{sec:photo}

Given the projected positions of the emission knots within Minkowski's H~II region, which is now known to lie at the distance of Cas~A (see \S\ref{sec:DiscMink} above), we posit these emission knots are CSM material from the progenitor. The spectra of these nebulae are unlike that expected from photoionization due to a blackbody. That is, they exhibit unusually strong [\ion{S}{2}] emission like that seen in SNR shocks, along with varying strengths of [\ion{O}{3}] and [\ion{S}{3}] emission, but no appreciable [\ion{O}{1}] or [\ion{O}{2}] emissions.

Below we investigate three possible excitation mechanisms of these nebulae: 1) flash photoionization by EUV emission during shock breakout \citep[e.g.][]{vandenBergh71a,CK78,chevalier05}, 2) photoionization by EUV and X-ray emission from the Cas~A remnant itself \citep{CK78}, or 3) shock emission from interaction with the blast wave or fast-moving knots.

Flash photoionization by the EUV emission during the SN event is an appealing explanation for the emission as it would be analogous to the early emission seen from the CSM ring around SN~1987A \citep{Lundqvist91,Lundqvist96}. However, SN~1987A was classified as Type IIpec with a blue supergiant progenitor, while Cas~A was classified as Type IIb with a likely red or yellow supergiant progenitor. Since both progenitors lost much of their outer envelopes, a first guess is that the EUV spectral shapes might have been similar, but the luminosities could be substantially different. \citet{Matzner99} find that the UV flash from a red supergiant should have an effective temperature near $5 \times 10^5\unit{K}$ and an energy of about $2 \times 10^{48}\unit{erg}$, which is roughly 20 times the energy they predict for a blue supergiant and that assumed by \citet{Lundqvist96}. While the Cas~A clouds are some 35 times farther from the explosion than the SN~1987A ring, the electron densities are about 100 times smaller, so a 10 times brighter EUV burst would give a similar ionization structure. 

\citet{Lundqvist96} found that the inner edge of the illuminated gas could be heated to temperatures around $10^6\unit{K}$.  We explored this emission with the current version of the shock spectrum code of \citet{Raymond79} and \citet{Cox85}. We varied the post-shock temperature, set the magnetic field to a high enough value to prevent compression as the gas cools, and turned off photoionization of the cooling gas by the hotter gas that would be upstream in a post-shock flow. Adopting a current remnant age of 350 yr and taking into account the light travel time from the SN event to the clouds about 25 light years from the explosion, the clouds would have been illuminated about 325 years ago. We then looked at the predicted emissivity after the gas has cooled and recombined for 325 years. Here we have assumed near solar abundances, though it is plausible that He and N could be somewhat enhanced, but far less so than that observed in the QSFs.

If the gas is initially heated to about log T = 5.7, the emission after 325 years is in qualitative agreement with the observed spectra. For densities between 200 and 400$\unit{cm}^{-3}$, the gas produces fairly strong [\ion{N}{2}] and [\ion{S}{2}] emission (4--24\% and 47--91\% compared to H$\alpha$), with undetectable \ion{He}{1}, [\ion{O}{2}] and [\ion{O}{3}] lines and weak [\ion{O}{1}] (${\sim}$10\% of H$\alpha$).  A density of 100$\unit{cm}^{-3}$ gives [\ion{O}{3}] and [\ion{S}{3}] emission of 97\% and 35\% of H$\alpha$, respectively, while densities around 75$\unit{cm}^{-3}$ give [\ion{O}{3}] several times brighter than H$\alpha$ as seen at position E3 in Table~\ref{table:linemeas} (after reddening correction).

Hence, it appears possible to match the observed spectra of these surrounding CSM nebulosities reasonably well with a UV flash from the Cas~A supernova. However, the range of initial temperatures that provide a match is somewhat narrow, in that for initial temperatures above $10^6$\,K gas at the observed densities does not cool enough to produce the observed emission lines, while for initial temperatures below about $4\times 10^5$\,K, the gas cools and recombines to the point that the forbidden lines are weak and the electron densities fall below 400$\unit{cm}^{-3}$ because the gas recombines. This could perhaps be understood as a selection effect, in that neither overionized nor underionized gas is readily observable at visible wavelengths. 

Next we investigated the continued photoionization of the gas by X-ray and EUV emission from the SNR itself. The remnant's current X-ray luminosity of $5\times 10^{37} \unit{erg}\unit{s}^{-1}$ \citep{Decenko2010} implies an ionization parameter  $\xi = L/nr^2$ much less than 1 in the clouds, which would not produce the observed ionization state or temperature \citep{Kallman82}. Ionization by photons produced in the thin ionization zone behind the reverse shock \citep{Hamilton88} would produce an ionization time of 150--1500 years at the distance of the clouds from the SNR, which is closer, but still not adequate to explain the fully ionized gas we observe. The EUV emission from the optical knots is not known but it is likely to be somewhat smaller based on the [\ion{O}{3}] optical flux of \citet{Bevan2017} and shock models such as those of \citet{Blair2000}.

Finally, we considered the possibility that shock waves excite the emission. While most nebulosities lie outside the forward blast wave, some high-velocity ejecta knots have reached the region of the East Cloud (see Fig.~\ref{fig:disEast}). Each of these ejecta knots would produce bow shocks in the CSM gas, perhaps merging into a large scale shock when they overlap \citep{Hartigan2016}. Although speculative, such a shock wave complex might help to explain the high [\ion{O}{3}]/H$\alpha$ ratio for position E3, but this would require far more ejecta knots than visible in the {\sl{HST}} images.

In summary, our analysis suggests that the observed emission comes mostly from gas that is cooling and recombining after photoionization by the EUV flash that occurred at shock breakout. However, ejecta driven shock waves may contribute for clouds nearest to the remnant. Detailed models like those of \citet{Lundqvist96} are needed for a more complete investigation.

\subsection{Mass Estimate of Surrounding Nebulae}

Assuming our conclusion that the nebulosities inside Minkowski's H~II region are at the distance of Cas~A, we estimated the mass contained within these nebulae by assuming that East Cloud is representative of the other bright emission clumps. This appears likely since the morphological and spectroscopic properties are similar between the East Cloud and emission clouds to the north of Cas~A. 

H$\alpha$ imaging of the East Cloud shows that the strongest emission is contained within a linear radius of $0.35\unit{pc}$. The average electron density in the East Cloud is about  $\rm{n_e}\approx100\unit{cm}^{-3}$ (see Table~\ref{table:linemeas}). We then assumed that this region can be roughly described by a sphere, with a filling factor of 0.3, so that only 30 percent of the sphere is filled with material with 1.1 hydrogen atoms per electron.

From these assumptions, we estimate that the brightest region of the East Cloud has a mass of $\approx 0.1\unit{M_\sun}$. Taking this mass estimate as the characteristic mass for the brightest emission nebulae surrounding Cas~A, we find that the total mass of these relatively dense clumps of emission detected in our optical images around Cas~A is $\approx1.5\unit{M_\sun}$. Since this mass estimate only accounts for the brightest optically detected emission, Cas~A's progenitor would have likely lost significantly more mass than what we have detected.

\subsection{Curious Western Emission `Streaks'}

The detection of numerous parallel emission filaments or `streaks' west of the remnant was quite unexpected (see Fig.~\ref{fig:Streaks}). Although their morphology might give the initial impression of wind-blown features, this is unlikely. Despite a few being in rough alignment with Cas~A's expansion center, this is of little importance given their strong non-convergence arrangement and their presence across such an extended north-south region. Their near parallel alignment makes equally unlikely an origin reflective of the progenitor pre-supernova mass-loss history mapped onto its proper motion. 

If the emission streaks were to lie at the distance of Cas~A and are CSM, their morphology might be indicating an unusual local magnetic field made visible by a similar ionization source as other CSM (see \S\ref{sec:photo} above). In any case, we note that the streaks do lie slightly west of a dust lane separating them from the northern CSM emission clumps.

Without kinematic or spectral data, the nature of this rather remarkable parallel set of emission streaks is unclear, as is how might they have an origin related to any of the other emission features projected near Cas~A. Thus, at present, they remain an interesting curiosity worthy of further investigation.


\subsection{Cas~A's Progenitor Mass-Loss History}

Due to the presence of dense clumps of circumstellar material in the form of N-rich QSFs, it has been postulated since the 1960's that Cas~A's progenitor experienced significant mass loss prior to it's explosion. Proper motions studies suggest that this mass-loss episode took place up until at least 11,000 years prior to explosion \citep{KampervandenBergh76}, although this estimate does not take into account acceleration of the QSFs due to interaction with the remnant's $\sim5000\unit{km}\unit{s}^{-1}$ forward shock. While most QSFs are found within a two arcminute radius centered on the center of expansion, they are not distributed in a spherical pattern, with a significant extension toward the southwest (see Fig.~\ref{fig:MinkowskiHII}), consistent with the near-IR QSF study by \citet{Koo2018}. 

The stark composition differences between the remnant's main shell of hydrogen-poor ejecta knots and the slower hydrogen-rich QSFs is strong evidence that Cas~A's progenitor lost much of its hydrogen-rich material before it exploded \citep{KampervandenBergh76,CK78}. This view is supported by the especially high-velocity, outer ejecta knots ($8000$--$12000\unit{km}\unit{s}^{-1}$) exhibiting [\ion{N}{2}]/H$\alpha$ ratios as high as 30 \citep{FBB87,FB91,Fesen01}, far in excess of that seen in the already N-rich QSFs, which indicate that the Cas~A progenitor had only a thin layer of photospheric hydrogen-rich material remaining at time of explosion. This conclusion is further supported by light echo spectra which showed the Cas~A SN was a likely SN~IIb event \citep{Krause08,Rest08,Rest11}.

However, several questions remain about Cas~A's progenitor's mass-loss history. The early evolution of a supernova remnant depends strongly on the progenitor's mass-loss history as its expanding ejecta and shock front encounter surrounding CSM \citep{Patnaude17}. The expansion of Cas~A has been generally modeled as an interaction between the ejecta and CSM formed by a dense stellar wind from a red supergiant (RSG) with a mass loss rate of $2\times10^{-5}\unit{M_\sun}\unit{yr}^{-1}$ and wind velocity of $10 \unit{km}\unit{s}^{-1}$ \citep{CO03}. While these models can be successfully applied to the kinematics of the blastwave, they are only applicable to the current evolutionary state of the remnant and are not useful in describing the new observations presented here. In light of this, here we explore the progenitor evolution starting from the main sequence which, to our knowledge, has not been previously considered.

Our H$\alpha$ imaging shows Cas~A's CSM lies between approximately 5 and 15$\unit{pc}$ from the center of expansion (see Figs.~\ref{fig:MinkowskiHII}--\ref{fig:Streaks}). The location of these optical features combined with the extended H~II region to the north provides clues as to the spectral classification of the Cas~A progenitor. Both the ionizing radiation from the progenitor, as well as the energy deposited in the environment via the stellar wind must be accounted for. 

\citet{mckee84} considered the evolution of photoionized bubbles around massive stars. In a homogeneous environment, the Stromgren radius is given as $\rm{R_{St}} = 66.9(S_{49}/\rm{n_{amb}})^{1/3}\unit{pc}$, where $\rm{S_{49}}$ is the Ly-$\alpha$ ionizing radiation in units of $10^{49}\unit{s}^{-1}$ and $\rm{n_{amb}}$ is the ambient medium density. Little is known about the pre-main sequence environment of Cas~A, so we assume that it evolved in the warm ISM with $\rm{n_{amb}} \simeq 3\unit{cm}^{-3}$. 

Numerical models and studies of the X-ray emitting ejecta in Cas~A point to $\sim3\unit{M_\sun}$ of shocked material, with $<0.5\unit{M_\sun}$ of unshocked material \citep{HwangLaming12, Young06}. Currently, Cas~A is expanding into a circumstellar environment with $\rm{n_{amb}}\approx 1\unit{cm}^{-3}$ at the forward shock radius, $\rm{R_{FS}}$, of $2.5\unit{pc}$ \citep{LPHS14}. For RSG models with isotropic mass loss, this corresponds to a mass-loss rate of $\dot{M}\lesssim2\times10^{-5}\unit{M_\sun}\unit{yr}^{-1}$. Integrating over the expected lifetime of a RSG (${\sim10^{5-6}\unit{yr}}$) yields a mass of material in the circumstellar environment of $\lesssim10 \unit{M_\sun}$. When combined with the amount of shocked and unshocked ejecta and the compact remnant mass, we estimate a main sequence progenitor mass of $\sim15\unit{M_\sun}$. The estimated
$\lesssim$ 10 M$_{\sun}$ of circumstellar material is broadly consistent with mass loss estimates
from evolutionary models of $15\unit{M_\sun}$ stars \citep{Georgy2015}.

A star's main sequence mass determines its spectral type, its ionizing flux, and stellar mass-loss. \citet{abbott82} lists stars with a zero-age-main-sequence (ZAMS) mass of $\sim15\unit{M_\sun}$ as likely B0.5~V stars, with main sequence lifetimes of $11.1\unit{Myr}$ and total mechanical wind energy injected into the interstellar medium of $8.5\times 10^{48}\unit{erg}$. This corresponds to an average wind luminosity, $L_w$ of $2.4\times10^{34}\unit{erg}\unit{s}^{-1}$. For traditional wind velocities of early type B stars, this corresponds to a mass loss rate of $\gtrsim10^{-7}\unit{M_\sun}\unit{yr}^{-1}$. 

According to \citet{panagia73}, O and B type stars are a significant source of ionizing radiation, which can photoevaporate a region to $<100\unit{pc}$, depending upon spectral type \citep{mckee84}. Thus, the mechanical wind expands into a photoevaporated and homogenized medium. As noted above, we assume a typical ``warm" ISM value and all values scale relative to this. From \citet{panagia73}, the ionizing luminosity for a B0  star is $\approx10^{48}\unit{s}^{-1}$ and $\approx10^{47}\unit{s}^{-1}$ for a B$0.5$ star. The Stromgren radii for stars of these spectral types are therefore between $\sim7$ and $15\unit{pc}$. The location of the inner edge of the H~II region scales as $\rm{n_{amb}}^{-1/3}$, so lower density interstellar environments will result in larger Stromgren radii, for a given spectral type. 

The wind of a main sequence star expands into this homogenized environment, evacuating a cavity \citep[c.f.][]{castor75}. The radius of the cavity evolves in a self similar manner $\propto L_w^{1/5}t^{3/5}$, where $t$ is the age of the wind. Using mass loss rates, wind velocities, and typical ages for O- and B-type stars, \citet{chevalier99} estimated radii for wind blown bubbles of $\sim5$--$10\unit{pc}$. He noted a large difference in the Stromgren sphere radius between B0~V and B1~V stars, consistent with the ionizing luminosities cited by \citet{panagia73}. 

Based on our observations showing the presence of diffuse extended emission at $\gtrsim10\unit{pc}$ away from Cas~A, we can rule out stars with spectral type later than B0 as main sequence progenitors. However, according to \citet{chevalier99}, the radius of the ionized shell is comparable to the wind blown bubble radius, $\sim10\unit{pc}$. By comparison, late type O (O9~V) stars will produce similar bubbles to early B0 stars, so it's possible that the Cas~A progenitor could have been a late O star, within a limited mass range.

For massive stars, the main sequence phase is followed by the red supergiant phase, which is dominated by slow, dense winds, with velocities of $\sim10$--$20\unit{km}\unit{s}^{-1}$ and mass loss rates of $\sim0.5$--$2.0\times10^{-5}\unit{M_\sun}\unit{yr}^{-1}$ \citep{smith14}. While there are currently large uncertainties about the mass loss rates of RSGs based on observations \citep{Georgy2017}, it is believed that most of the progenitor's mass is lost during the RSG phase, which lasts for $\simeq1.0\unit{Myr}$ \citep{Eldridge2018}. The RSG wind expands into the low density main sequence wind, eventually forming a wind-blown, radiatively cooled shell. This swept-up shell is expected to expand outward until the dynamical age is comparable to the cooling time of the swept-up material.

We can estimate the radius of the RSG shell by making assumptions on the mass loss rate and wind velocity, in order to set limits on the RSG phase. In their study of Cas~A's pre-SN wind, \citet{CO03} estimated a RSG wind speed of $10\unit{km}\unit{s}^{-1}$ and a mass loss rate of $\dot{M} = 2 \times 10^{-5}\unit{M_\sun}\unit{yr}^{-1}$. The RSG wind expanded into the wind sculpted circumstellar environment of the main sequence progenitor. The main sequence wind blown bubble density can be approximated using the self-similar expansion relations of \citet{castor75}. Assuming a main-sequence mass-loss rate of $\dot{M}\approx10^{-8}\unit{ M_\sun}\unit{yr}^{-1}$ and wind velocity of $\mathrm{v_{wind}} = 500\unit{km}\unit{s}^{-1}$, the density of the CSM at a radius of $1\unit{pc}$ is $<10^{-3}\unit{cm}^{-3}$. For simplicity, we assume that the density of the CSM left behind by the main sequence wind is $\sim5\times 10^{-26}\unit{gm}\unit{cm}^{-3}$. 

The radius of the RSG swept up shell evolves self-similarly \citep[e.g.,][]{weaver77}:

\begin{equation}
R(t) = 0.88 \left(\frac{L_w}{\rho_{\mathrm{amb}}}\right)^{1/5} t^{3/5}
\end{equation}

\noindent
where $t$ is the age of the RSG wind. We can estimate $t$ in the above equation by assuming a $15\unit{M_{\sun}}$ progenitor, and accounting for the observed shocked ejecta and compact remnant mass. Given the $\lesssim 10\unit{M_\sun}$ of material we estimate is lost to the RSG wind, and the estimated mass loss rate of $2\times 10^{-5}\unit{M_\sun}\unit{yr}^{-1}$, we estimate that the RSG phase lasted for $\approx$ $5\times10^5\unit{yr}$. Therefore, the RSG sweeps up a shell out to a radius of $\sim10\unit{pc}$. This value is in good agreement with the features observed around Cas~A as shown in Figure~\ref{fig:MinkowskiHII}.

The QSF emission knots are dense clumps circumstellar material, with high nitrogen abundances suggesting their origin arises from a hydrogen-stripped envelope. The QSF precursor material was lost from the progenitor during approximately the last 200,000 years from the deep layers of the outer envelope. Previously suggested by \citet{CO03}, the QSFs observed properties can arise as the result of shocked over-densities in the smooth RSG wind.

Finally, we now considered the phase of evolution just prior to core collapse and explosion. Evidence for a post-RSG phase comes from {\it Chandra} observations of Cas~A's shocked metal-rich ejecta. \citet{HwangLaming09} analyzed shocked ejecta in Cas~A and compared the X-ray spectra to model spectra expected from the interaction between ejecta and an isotropic RSG wind. They found that, in order to match the line intensity ratios of H- to He-like ions such as silicon, they needed to insert a low density cavity around the Cas~A progenitor, with $r \lesssim0.2\unit{pc}$. They argued that the inclusion of the cavity allowed the ejecta to adiabatically expand and cool before running into the dense RSG wind, effectively delaying deposition of thermal energy into the ejecta which leads to the bright X-ray emission \citep{Patnaude17}.

\citet{Schure08} explored the effects of the circumstellar environment on the evolution of Cas~A's northeast jet. They argued that, given the energetics of the ejecta in the jet \citep{Fesen01,Laming06}, a small cavity formed by a high velocity wind could be accommodated, but if the cavity were too large, as a result of an extended evolutionary phase which blows the CSM away from the progenitor, then the jet would be buried in the CSM. In both \citet{HwangLaming09} and \citet{Schure08}, the predicted high-velocity wind phase is short, $<<$ 10,000 years. This leads to the question of what formed the cavity? A natural explanation would be for a post-RSG Wolf-Rayet (WR) phase. WR stars have mass loss rates similar to red supergiants, but with winds speeds two orders of magnitude higher. However, the low mass of the progenitor is incompatible with a WR phase \citep{Sravan19}. In addition, hydrodynamical simulations of \citet{vanVeelen09}, showed the QSFs cannot be explained by a WR wind and the remnant's emission did not show evidence for remainders of a WR shell.

While a WR progenitor for Cas~A seems unlikely, several stripped envelope SNe that exhibit progenitor properties similar to those estimated for Cas~A, including SN~1993J, SN~2011dh and SN~2013df, have suspected compact yellow supergiants (YSG) progenitors \citep{Smartt09,Bersten2012,Yoon2017}. Model calculations for $12$--$15 \unit{M_\sun}$ RSGs with enhanced mass loss during some part of their evolution have led to YSGs as possible progenitors of SNe IIb \citep{Georgy2012}. YSGs have typical mass-loss rates similar to RSGs but with wind speeds up to $20\times$ higher \citep{smith14}. They therefore have the required mechanical luminosity to clear out a small cavity around the progenitor. Additionally, \citet{Drout11} found that in M31 the YSG phase for $12$--$15\unit{M_\sun}$ ZAMS stars lasts ${\sim}3000$ years. Such a brief YSG, pre-SN evolutionary phase with a relatively high wind velocity could therefore account for the small circumstellar cavity around Cas~A which seems to be required by observations, yet remains compatible with its estimated progenitor mass.

In summary, our findings are consistent with a scenario where Cas~A's progenitor was initially a B0 or late type O star which photoionized and carved out a large circumstellar bubble of radius $\gtrsim 10\unit{pc}$. Then the star evolved as a red supergiant with a mass loss rate of $\sim 2\times 10^{-5}\unit{M_\sun}\unit{ yr}^{-1}$, for approximately $5\times10^{5}\unit{yr}$. During the latter stages of this phase, QSFs formed, due to localized over densities in the wind. Finally, in the last several thousand years, after much of the hydrogen-rich envelope was lost, the star evolved as a yellow supergiant. The winds of YSGs are substantially faster than those of RSGs, with similar mass loss rates, so the wind carved out a small cavity around the progenitor, which is compatible with the evolution of the northeast jet, and also can explain the ionization state of the X-ray emitting material.

We note that while a YSG progenitor for Cas~A is consistent with both these observations and existing data, other scenarios could result in a pre-SN cavity if one considers a binary progenitor system \citep{Pod1992,Aldering1994,Claeys2011,Folatelli2014,Ryder2018}. However, there is no evidence for a surviving companion in Cas~A \citep{Kochanek18,Kerzendorf2019}.

\section{Conclusions}
\label{sec:Con}

Although Cas~A is one of the best studied Galactic supernova remnants, the nature of faint surrounding emission nebulae, first described by Minkowski a half century ago, had not been fully investigated. Here we present an optical imaging and spectral study of the circumstellar environment around Cas~A extending out to a radius of roughly $20\unit{pc}$, at the remnant's assumed $3.4\unit{kpc}$ distance. Our main findings are summarized below.

1. We detected considerable clumpy H$\alpha$ emission approximately 10\arcmin\ to 20\arcmin\ to the northwest, north, and east of Cas~A which are morphologically and spectroscopically distinct from that of neighboring H~II region emission. We interpret this emission as RSG mass-loss and postulate that this material accumulated against the southern edge of extended local H~II region.

2. {\sl Spitzer} infrared images of Cas~A show that apparent breaks in the H$\alpha$ emission around the remnant are likely due to line-of-sight dust obscuration. This suggests a more continuous emission nebula may exist north and east of Cas~A. 

3. We present evidence that an emission nebulae east of Cas~A (the `East Cloud'), lies at Cas~A's distance due to direct interactions between this cloud and the remnant's high-velocity ejecta knots. Due to similar morphology and spectra of neighboring nebulae with the East Cloud, we conclude these nebulae also lie adjacent to Cas~A and represent progenitor mass-loss material. 

4. The optical spectra of the emission nebulae surrounding Cas~A exhibit unusually strong [\ion{S}{2}] relative to H$\alpha$, typically an indicator of shock heated gas even though the nebulae lie outside the forward shock. However, observed spectra were modeled as material which was photoionized by the EUV flash from the Cas~A supernova. 

5. Based on the observed electron densities between less than $100\unit{cm}^{-3}$ and $450\unit{cm}^{-3}$, we estimate that approximately 1.5 solar masses of material is contained within the denser emission clumps detected in our images.

6. Our findings suggest the progenitor first underwent considerable mass-loss from a fast main-sequence wind, then from a slower, clumpy RSG wind which generated the observed emission nebulae that partially surrounds the remnant. Finally, just prior to explosion, the progenitor likely underwent a brief high-velocity wind phase, like that from a YSG.  

In summary, our study of the circumstellar environment around Cas~A clarifies the nature of Minkowski's so-called `H~II region' as red supergiant mass-loss material from Cas~A's progenitor, photoionized by EUV emission during shock breakout. It also indicates that the circumstellar QSFs must have formed fairly close to the time of stellar outburst. Our findings support previous X-ray results that the forward shock-front is expanding into a non-uniform circumstellar medium formed during the progenitor's RSG phase \citep[e.g.,][]{Patnaude17}.

One by product of this research is a note of caution regarding extragalactic SNR studies. The fact that the clumpy RSG emission nebulae $10$ to $20\unit{pc}$ north of Cas~A exhibit [\ion{S}{2}]/H$\alpha$ line ratios like that typically observed in SNRs, it might when viewed at low spatial resolution be misinterpreted as a SNR. This could result in an incorrect interpretation of this emission as either a much larger SNR with a young O-rich remnant inside, or a young SNR neighboring an older one. 

We note that Cas~A-like SN~IIb events such as SN~1993J and SN~2011dh which remain relatively bright many years post-max may not be due to interaction with an isotropic RSG mass-loss material as often been proposed, but rather closer-in and dense material, similar to material that formed Cas~A's QSFs. In this way, Cas~A's mass-loss history may be a useful guide for studies of extragalactic SN~IIb events, and in models of the late stages of high-mass stellar evolution.

\acknowledgments{
We thank Bon-Chul Koo for useful conversations, and Eric Galayda and the MDM and MMT staffs for assistance with our observations. K.E.W.\ and R.A.F.\ acknowledge support from STScI Guest Observer Programs 15337 and 15515. 
K.E.W.\ also acknowledges support from Dartmouth's Guarini School of Graduate and Advanced Studies, and the Chandra X-ray Center under CXC grant GO7-18050X. 
D.J.P.\ acknowledges support from NASA contract NAS8-03060.
R.A.C.\ acknowledges support from NSF grant 1814910.
This research is based in part on observations made with the NASA/ESA Hubble Space Telescope obtained from the Space Telescope Science Institute, which is operated by the Association of Universities for Research in Astronomy, Inc., under NASA contract NAS5-6555. 
Some of the observations reported here were obtained at the MMT Observatory, a joint facility of the Smithsonian Institution and the University of Arizona. 
This work is also based in part on observations made with the Spitzer Space Telescope, which is operated by the Jet Propulsion Laboratory, California Institute of Technology under a contract with NASA. 
Finally, we made use of the software package Montage which was funded by the National Science Foundation under Grant Number ACI-1440620, and was previously funded by the National Aeronautics and Space Administration's Earth Science Technology Office, Computation Technologies Project, under Cooperative Agreement Number NCC5-626 between NASA and the California Institute of Technology.}

\facilities{Hiltner (OSMOS, ModSpec), McGraw-Hill (LBNL imaging CCD), McD:0.8m (Prime Focus Corrector), MMT (Binospec), Spitzer (IRAC), HST (ACS)}

\software{PYRAF \citep{PYRAFcite}, AstroPy \citep{AstropyCiteA,AstropyCiteB}, ds9 \citep{ds9Cite}, Binospec \citep{BinospecCite}, Montage \citep{Montagecite}, L.A.\ Cosmic \citep{vanDokkum01}, Raymond Shock Code \citep{Raymond79,Cox85} }

\bibliography{bibmaster.bib}

\end{document}